# Xona Pulsar Compatibility with Spaceborne GNSS Receivers

Argyris Kriezis, Max Turner, Claire Mah, Michael O'Meara, & Tyler G. R. Reid

*Xona Space Systems*

**ABSTRACT**

Low Earth Orbit (LEO) Positioning, Navigation, and Timing (PNT) systems operating in L-band Radionavigation-Satellite Service (RNSS) spectrum are rapidly emerging through both commercial and governmental initiatives. These systems are intended to complement traditional Medium Earth Orbit (MEO) constellations, such as GPS and Galileo, by providing improved satellite geometry, higher received signal power, and greater orbital diversity. As with any new RNSS deployment, ensuring compatibility with existing systems is essential to prevent harmful interference and preserve established services. While compatibility assessments for terrestrial GNSS receivers have been studied extensively, compatibility with spaceborne receivers has received comparatively little attention due to the large separation distances between MEO GNSS constellations and most spacecraft. In contrast, LEO PNT systems operate in close proximity to other satellites in LEO, making spaceborne receiver compatibility an important consideration. This paper extends the established ITU-R compatibility assessment methodology, originally developed for terrestrial receivers, to spaceborne GNSS users. The analysis evaluates the factors affecting carrier-to-noise ratio ($C/N_0$) degradation of legacy GNSS signals in orbital environments. Results show that, although spaceborne receivers may experience higher received power levels from nearby LEO PNT satellites, such as Xona's Pulsar constellation, they can operate compatibly alongside legacy RNSS.

## 1. INTRODUCTION

More than two hundred navigation satellites have been launched into Low Earth Orbit (LEO) in recent years, with thousands more proposed by over ten emerging providers. These systems employ a wide range of frequency bands, including traditional RNSS L-band allocations as well as C-, S-, UHF-, and VHF-band spectrum. For systems operating in RNSS frequencies, compatibility with existing Medium Earth Orbit (MEO) and Geostationary Orbit (GSO) navigation systems is critical to ensuring the continued reliability of navigation services. Table 1 summarizes a selection of these emerging LEO PNT systems.

Among these initiatives, Xona is deploying Pulsar, a commercial LEO PNT system consisting of approximately 260 satellites operating in dual L-band frequencies adjacent to GPS L1 and L5. Pulsar is designed to be compatible and interoperable with existing GNSS while providing additional capabilities, including stronger received signals, centimeter-level orbit and clock products, encrypted services, and range authentication. In 2025, Xona demonstrated that Pulsar signals, despite being transmitted at power levels up to two orders of magnitude higher than GPS signals, do not cause harmful interference to legacy GNSS services (Reid, et al., 2025). That assessment combined theoretical analysis based on ITU-R M.1831 with laboratory testing of commercial GNSS receivers and live-sky measurements using the in-orbit Pulsar-0 satellite.

*Table 1: Comparison of dedicated LEO PNT systems, deployments, and plans. Note that satellites already deployed were verified on celestrack.org.*

| **Constellation** | **Flag** | **Status** | **Frequency Band(s)** | **Launches** | **Constellation Size** | **Ref.** |
|---|---|---|---|---|---|---|
| Iridium STL | USA | Active | L (non-RNSS) | 2017 (NEXT) | 66 (active) | (Riley, 2023) |
| Xona Pulsar | USA | Deploying | L | 2022 (demo)<br>2025 (ops)<br>2026+ (planned) | 2 (launched)<br>258 (planned) | - |
| TrustPoint | USA | Demo / Planned | C | 2023, 2025 (demo)<br>2026+ (planned) | 3 (launched)<br>288 (planned) | (Anderson, 2023)<br>(Khalil, 2025) |
| ESA LEO-PNT | Europe | Demo / Planned | L, S, C, UHF | 2025-2028 (demo) | 2 (launched)<br>8 (demo, planned)<br>263 (planned) | (FrontierSI, 2024)<br>(Sarnadas, 2025) |
| JAXA | Japan | Feasibility | C | 2030 (P1 planned)<br>2035 (P2 planned) | 240 (P1 planned)<br>480 (P2 planned) | (Rubinov, 2024)<br>(Elena, 2025) |
| ArkEdge Space | Japan | Feasibility | VHF | - | 50-100 (planned) | (Rubinov, 2024) |
| Fergani Space | Turkey | Demo / Planned | L, S, Ku, Ka | 2025 (demo)<br>2026+ (planned) | 1 (launched)<br>120 (planned) | (FrontierSI, 2024)<br>(Newsroom, 2024) |
| CentiSpace | China | Deploying | L | 2018 - 2022 (5 demos)<br>2025+ (10 ops) | 26 (launched)<br>190 (planned) | (Rubinov, 2024)<br>(Xucheng, 2023) |
| Geely (Geespace) | China | Deploying | L, S, Ku, Ka | 2022 - 2024 (ops)<br>2025+ (ops) | 25* (launched)<br>240 (planned) | (Rubinov, 2024) |
| SatNet LEO (Hulianwang) | China | Deploying | L | 2023 – 2025 (ops)<br>2025 (planned)<br>2030 (planned) | 168 (launched)<br>504 (Final planned) | (Rubinov, 2024)<br>(Guowang, 2025) |
| GNSSaS | UAE | Planned | L, S | - | demo planned | (FrontierSI, 2024) |
| VyomIC | India | Planned | - | - | 125 – 150 (planned) | (GC NewsDesk, 2025) |

*Although 25 Geespace satellites have been launched, it is unclear to the authors if these are part of the dedicated 240 satellites as part of the PNT sub-constellation. The larger Geely constellation has been announced to be 5,676 satellites (Rubinov, 2024).

Historically, compatibility assessments have focused on terrestrial and airborne receivers. However, satellites in Earth orbit also rely extensively on GNSS for navigation, timing, and orbit determination. Coexistence among traditional MEO constellations, such as GPS and Galileo, has not posed significant challenges because these systems operate at altitudes of approximately 20,000 km and are therefore spatially separated from most spacecraft. In contrast, LEO PNT systems operating at altitudes of approximately 800–1200 km are deployed in regions populated by numerous active satellites. The reduced separation distance can result in higher received signal power levels at nearby spaceborne GNSS receivers than would be experienced by terrestrial users, potentially increasing the effective noise floor. This effect is partially mitigated by the smaller number of LEO PNT satellites simultaneously visible from another spacecraft, which reduces aggregate received power.

This paper presents a compatibility assessment of the Xona Pulsar system with GNSS receivers operating aboard satellites at altitudes between 400 km and 1000 km. Compatibility is evaluated in terms of $C/N_0$ degradation of GPS and Galileo resulting from the introduction of Pulsar transmissions. Specifically, the analysis considers the impact of the Pulsar X1 signal on the adjacent GPS L1/Galileo E1 band and the Pulsar X5 signal on the neighbouring GPS L5/Galileo E5 band. In the absence of a formal methodology for evaluating compatibility with spaceborne receivers, the terrestrial framework defined in ITU-R M.1831 is adapted to orbital environments.

This work represents one of the first efforts to extend established RNSS compatibility methodologies to spaceborne users, building upon the methodology proposed by M. Paonni et al. at the 2025 ION GNSS+ Conference (Paonni, 2025). Although the analysis focuses on the Xona Pulsar constellation, the proposed framework is broadly applicable to other emerging LEO PNT systems and contributes to ensuring the compatible operation of next-generation navigation constellations alongside legacy GNSS services in an increasingly congested orbital environment.

## 2. METHODOLOGY

The ITU-R M.1831-1 document "A coordination methodology for radionavigation-satellite service inter-system interference estimation" outlines the methodology for evaluating the $C/N_0$ degradation for pre-existing systems when a new alternate system is introduced. The foundation of the method is that the $C/N_0$ degradation of a signal is the ratio of the new effective noise ($I_{alt}$) compared to the pre-existing effective noise. As presented in Equation 1, the pre-existing noise is the sum of four types of noise sources, the ambient thermal noise ($N_o$), the noise by adjacent-band non-RNSS system ($I_{ext}$), the noise of the RNSS system to itself ($I_{ref}$) and the noise of other RNSS system currently in operation ($I_{rem}$).

$$\Delta\left(\frac{C}{N'_{\text{o}}}\right) = 1 + \frac{I_{alt}}{N_{\text{o}} + I_{ext} + I_{ref} + I_{rem}} \quad (1)$$

where:

- $C$ Carrier strength of the impacted RNSS signals (W)
- $N_0$ Receiver thermal noise power spectral density (W/Hz)
- $I_{ext}$ Effective white noise power spectral density (W/Hz) due to non-RNSS "external" signal interference
- $I_{ref}$ Effective white noise power spectral density (W/Hz) due to interference from all the signals of the impacted "reference" RNSS systems to itself
- $I_{rem}$ Effective white noise power spectral density (W/Hz) due to interference from all existing "remaining" RNSS systems
- $I_{alt}$ Effective white noise power spectral density (W/Hz) due to interference from the new "alternate" RNSS system being introduced

The $N_o$ and $I_{ext}$ variables are relatively constant and are defined in the ITU-R M.1831-1 as -201.50 dB(W/Hz) and -206.5 dB(W/Hz). The remaining effective interference variables are a function of pre-existing and new RNSS signals, their power spectral density (PSD) and power level as seen in Equation 2.

$$I_{ref/rem/alt} = P_{max}^{R} + G^{agg} - L_{proc} + SSC \quad (3)$$

where:

- $P_{max}^{R}$ Maximum single satellite received power (W)
- $G^{agg}$ Aggregation gain factor for single worst-case satellite geometry (dB)
- $L_{proc}$ Processing loss
- $SSC$ Spectral Separation Coefficient (Between the impacted and impacting signals)

Consequently, the pre-existing effective noise environment differs among GNSS signals, resulting in different baseline interference levels for GPS L1 C/A, Galileo E1, QZSS, and other RNSS signals. A key parameter in this calculation is the Spectral Separation Coefficient (SSC), which converts received signal power into equivalent interference power by accounting for the spectral overlap between signals.

The aggregate gain term, ($G^{agg}$), represents the combined contribution of all satellites within a constellation. Because the geometry of a non-geostationary satellite orbit (NGSO) constellation varies continuously, ($G^{agg}$) changes with both time and user location. Under the M.1831-1 methodology, a conservative worst-case value is determined by evaluating the aggregate gain globally over a 24-hour period and selecting the maximum value observed. Although this approach may represent conditions experienced only briefly at a particular location, it provides a standardized basis for compatibility assessments and is widely used for terrestrial receiver evaluations.

The same general framework can be applied to spaceborne receivers, however, the geometry between transmitting and receiving satellites is considerably more dynamic than for terrestrial users. The received power from each RNSS signal must therefore be calculated as a function of time using free-space path loss (FSPL), transmit antenna gain, and receive antenna gain. As a result, both the pre-existing interference environment and the interference contribution from a new RNSS system vary continuously with orbital geometry.

Explicitly modeling the time-varying contributions of $I_{ref}$ and $I_{rem}$ for all operational RNSS constellations would substantially increase the complexity of the analysis while providing limited additional insight. Instead, a fixed and conservative pre-existing noise environment was assumed for all scenarios. A total pre-existing effective noise level of −200 dB(W/Hz) was adopted, corresponding to an aggregate pre-existing RNSS interference contribution of −211.7 dB(W/Hz) in addition to the $N_o$ and $I_{ext}$ terms defined in ITU-R M.1831-1. This assumption produces a lower noise floor than that encountered in most terrestrial compatibility analyses and therefore provides a conservative baseline for evaluating interference for a spaceborne receiver (Reid, et al., 2025).

*Table 2: Xona Pulsar input parameters.*

| Parameter | X1 | X5 |
| --- | --- | --- |
| Number of Satellites | 258 | |
| Satellite Altitude | 1080 km | |
| Inclination | 53 deg & 97.5 deg | |
| Center Frequency (MHz) | 1593.3225 | 1190.51625 |
| 99.5% Bandwidth (MHz) | 1.8 | 17.7 |
| Modulation | EFQPSK | EFQPSK |
| Chip Rate (Mcps) | 1.023 | 10.23 |
| Doppler Range (kHz) | 33.8 | 25.2 |
| Max Single Satellite RIP (dBW) [Ground] | -138.4 | -136.0 |
| GPS L1 C/A - Pulsar X1 SSC | -97.4 | - |
| GAL E1 BC - Pulsar X1 SSC | -86.2 | - |
| GPS L5 - Pulsar X5 SSC | - | -85 |
| GAL E5 - Pulsar X5 SSC | - | -86 |

The effective interference contribution of the proposed RNSS system, ($I_{alt}$), was evaluated using two complementary approaches. First, a static worst-case analysis was performed using a methodology analogous to that employed for terrestrial compatibility assessments. Second, a dynamic analysis was conducted that explicitly accounts for the relative motion of the LEO PNT satellites and the spaceborne receiver. The input parameters used for the Xona Pulsar constellation are summarized in Table 2.

## 3. COMPATIBILITY – STATIC WORST-CASE ANALYSIS

For the static analysis, the worst-case aggregate gain ($G^{agg}$) was calculated for a range of potential spaceborne receiver altitudes. Since the Pulsar constellation operates at approximately 1080 km altitude, receiver altitudes between 400 km and 1000 km were evaluated in 100 km increments. As receiver altitude increases, the received power from the Pulsar constellation also increases due to the reduced separation distance. However, the aggregate gain decreases because fewer Pulsar satellites are simultaneously visible.

Table 3 summarizes the maximum received power and corresponding worst-case aggregate gain values for each receiver altitude. The aggregate gain was determined using a global grid spanning latitudes from −90° to 90° and longitudes from −180° to 180° with a 1° resolution. The calculations were performed using the nominal Pulsar constellation geometry, anticipated X1 and X5 transmit antenna gain patterns, and the DO-235C/DO-292A aviation reference receiver antenna model.

Using the parameters from Table 3 and the assumed pre-existing effective noise level of −200 dB(W/Hz), the expected $C/N_0$ degradation was calculated in accordance with the ITU-R M.1831-1 methodology. Table 4 presents the resulting effective interference levels ($I_{alt}$) and corresponding $C/N_0$ degradations for each receiver altitude. Figure 1 resents the resulting worst-case $C/N_0$ degradation value for each legacy GNSS signal at each altitude.

*Table 3: Xona Pulsar max RIP and Gagg for each altitude.*

| Pulsar Altitude (km) | User Altitude (km) | X1 Maximum RIP (dBW) | X1 Aggregate Gain Factor | X5 Maximum RIP (dBW) | X5 Aggregate Gain Factor |
|---|---|---|---|---|---|
| 1080 | 400 | -134.46 | 7.19 | -131.84 | 7.42 |
| | 500 | -133.03 | 6.74 | -130.40 | 7.07 |
| | 600 | -131.33 | 6.25 | -128.68 | 6.51 |
| | 700 | -129.21 | 5.62 | -126.54 | 5.41 |
| | 800 | -126.39 | 5.01 | -123.71 | 4.75 |
| | 900 | -122.21 | 4.54 | -119.51 | 4.35 |
| | 1000 | -113.84 | 3.65 | -111.13 | 3.47 |

*Table 4: Calculated $I_{alt}$ values for each altitude.*

| Spaceborne Receiver Altitude (km) | GPS L1 C/A Ialt [dB(W/Hz)] | GAL E1 BC Ialt [dB(W/Hz)] | GPS L5 Ialt [dB(W/Hz)] | GAL E5 Ialt [dB(W/Hz)] |
|---|---|---|---|---|
| 400km | -225.7 | -214.5 | -210.5 | -211.4 |
| 500km | -224.7 | -213.5 | -209.4 | -210.3 |
| 600km | -223.5 | -212.3 | -208.2 | -209.2 |
| 700km | -222.0 | -210.8 | -207.2 | -208.1 |
| 800km | -219.8 | -208.6 | -205.0 | -205.9 |
| 900km | -216.1 | -204.9 | -201.2 | -202.1 |
| 1000km | -208.6 | -197.4 | -193.7 | -194.6 |

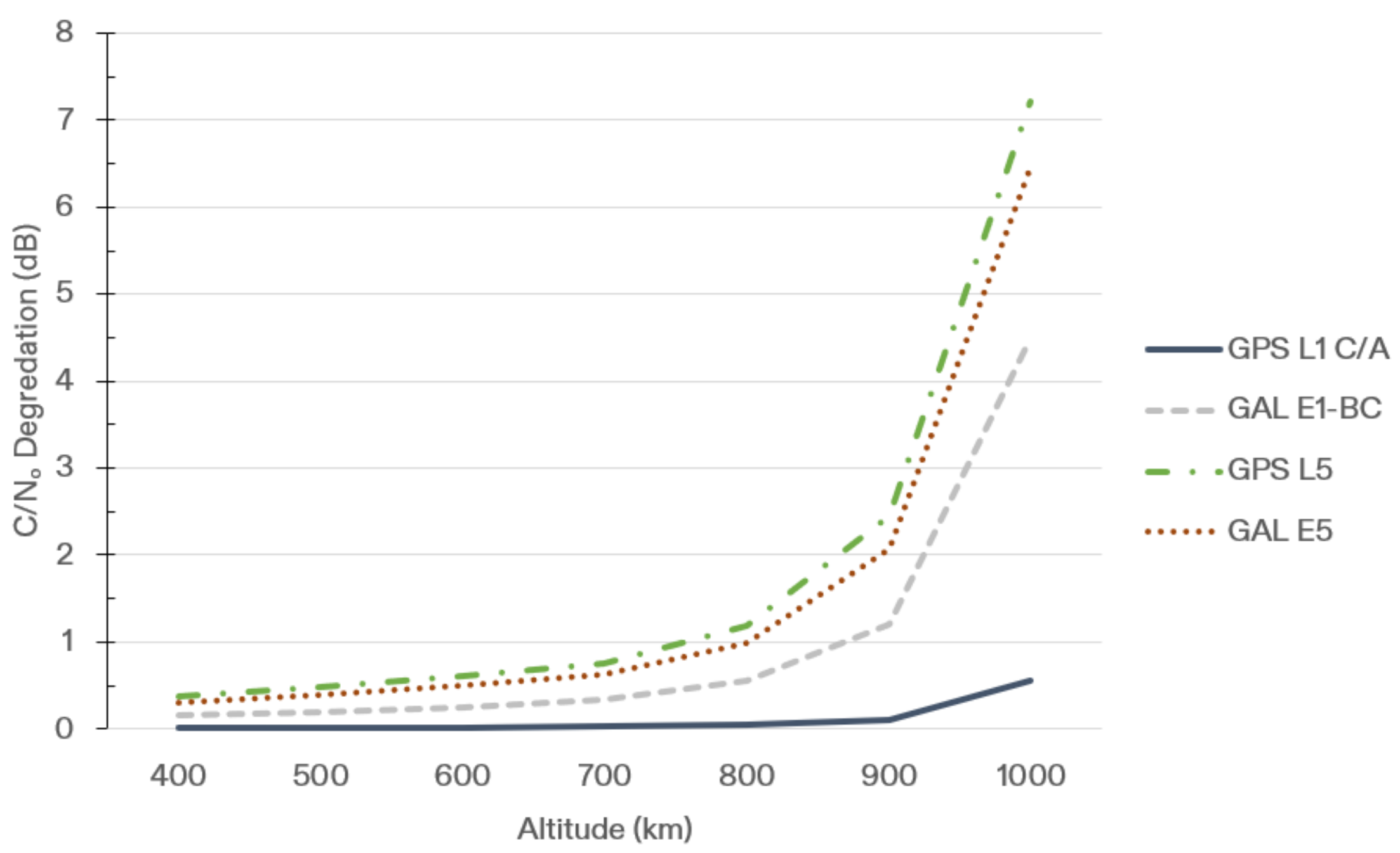

*Figure 1: Expected worst-case $C/N_0$ degradation at each altitude for GPS and Galileo.*

The results indicate that $C/N_0$ degradation remains limited for receiver altitudes between 400 km and 900 km, corresponding to separation distances of approximately 680 km to 180 km from the Pulsar constellation. At altitudes between 900 km and 1000 km, the calculated worst-case degradation can become more significant for some GNSS signals as the separation distance decreases. However, under these conditions, the temporal characteristics of the interference become increasingly important, since close-proximity geometries occur only for limited periods of time. This behavior is not unique to the Pulsar constellation and is generally applicable to any LEO PNT system. At sufficiently small separation distances, received signal power can become significant even for systems transmitting at lower power levels.

A limitation of the static analysis is that it represents only a worst-case snapshot and provides no information regarding the frequency or duration of the predicted degradation. These temporal characteristics become particularly important when two satellites operate in close proximity, as the time spent in such geometries is often brief. To address this limitation, a dynamic compatibility analysis was performed and is presented in the following section.

## 4. COMPATIBILITY – DYNAMIC ANALYSIS

While the static analysis provides a conservative upper bound on potential interference, it does not capture the temporal behavior of the degradation. A dynamic analysis was therefore performed to evaluate the distribution of $C/N_0$ degradation over time and to determine how frequently worst-case conditions occur. Rather than using a fixed worst-case $G^{agg}$, the aggregate gain was recalculated at each simulation time step based on the instantaneous geometry between the Pulsar constellation and the spaceborne receiver.

Because the static analysis showed limited $C/N_0$ degradation for receiver altitudes up to 900 km, the dynamic assessment focused on the most challenging case, a user satellite operating in close proximity to Pulsar at an altitude of 1000 km. The full 258-satellite Pulsar constellation was propagated over a 48-hour period using a 1-second time step, together with a user satellite at 1000 km altitude. To evaluate the impact of orbital geometry, the analysis was repeated for five user satellite inclinations ranging from 0° to 98°.

At each simulation time step, the received power from every visible Pulsar satellite was recomputed using the instantaneous free-space path loss, transmit antenna gain, and receive antenna gain. These values were then used to calculate the effective interference contribution, ($I_{alt}$), and the resulting $C/N_0$ degradation. Figure 2 presents the results of the of the dynamic analysis of the $C/N_0$ degradation for GPS and Galileo signals.

The results show that the worst-case values identified in the static analysis rarely occurs for satellites operating in close proximity to the Pulsar constellation. Across all simulated scenarios, the 99th-percentile degradation remained below 3 dB, while the 90th-percentile degradation remained well below 1 dB. These results indicate that significant degradation events are infrequent and short in duration. Figure 3 presents the duration complementary cumulative distribution of interval duration in which the degradation exceeds 3 dB, for the GPS L5 signal. GPS L5 was chosen as a worst-case representative since it experiences the largest $C/N_0$ degradation.

The analysis indicates that compatibility is driven primarily by separation distance rather than transmit power alone. As the distance between two spacecraft decreases, the received power naturally increases regardless of the specific LEO PNT system being considered. Consequently, brief periods of elevated $C/N_0$ degradation may occur whenever satellites operate in close orbital proximity.

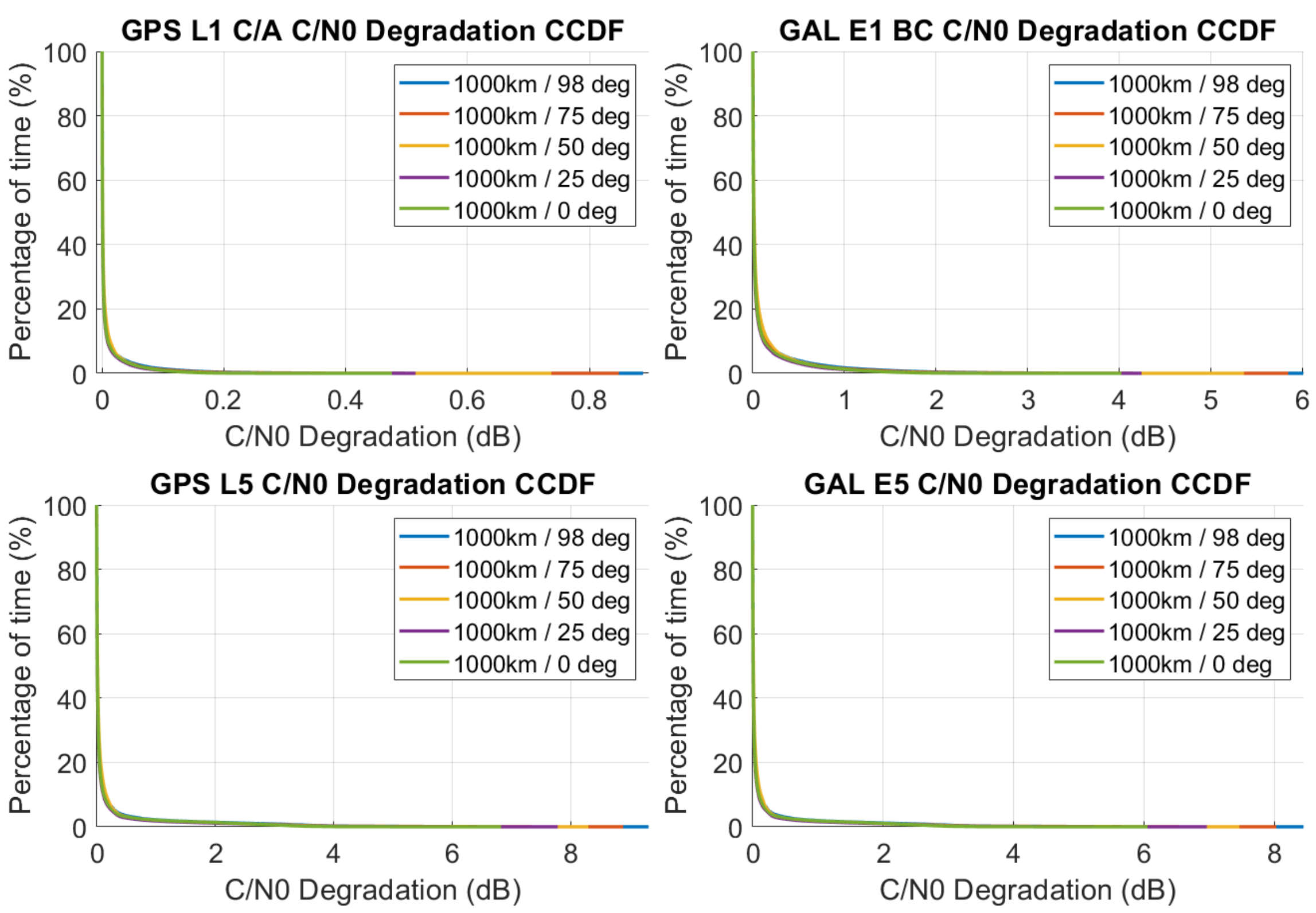

*Figure 2: Complementary cumulative distribution function (CCDF) of expected C/$N_0$ degradation from Pulsar for satellites at 1000km orbit.*

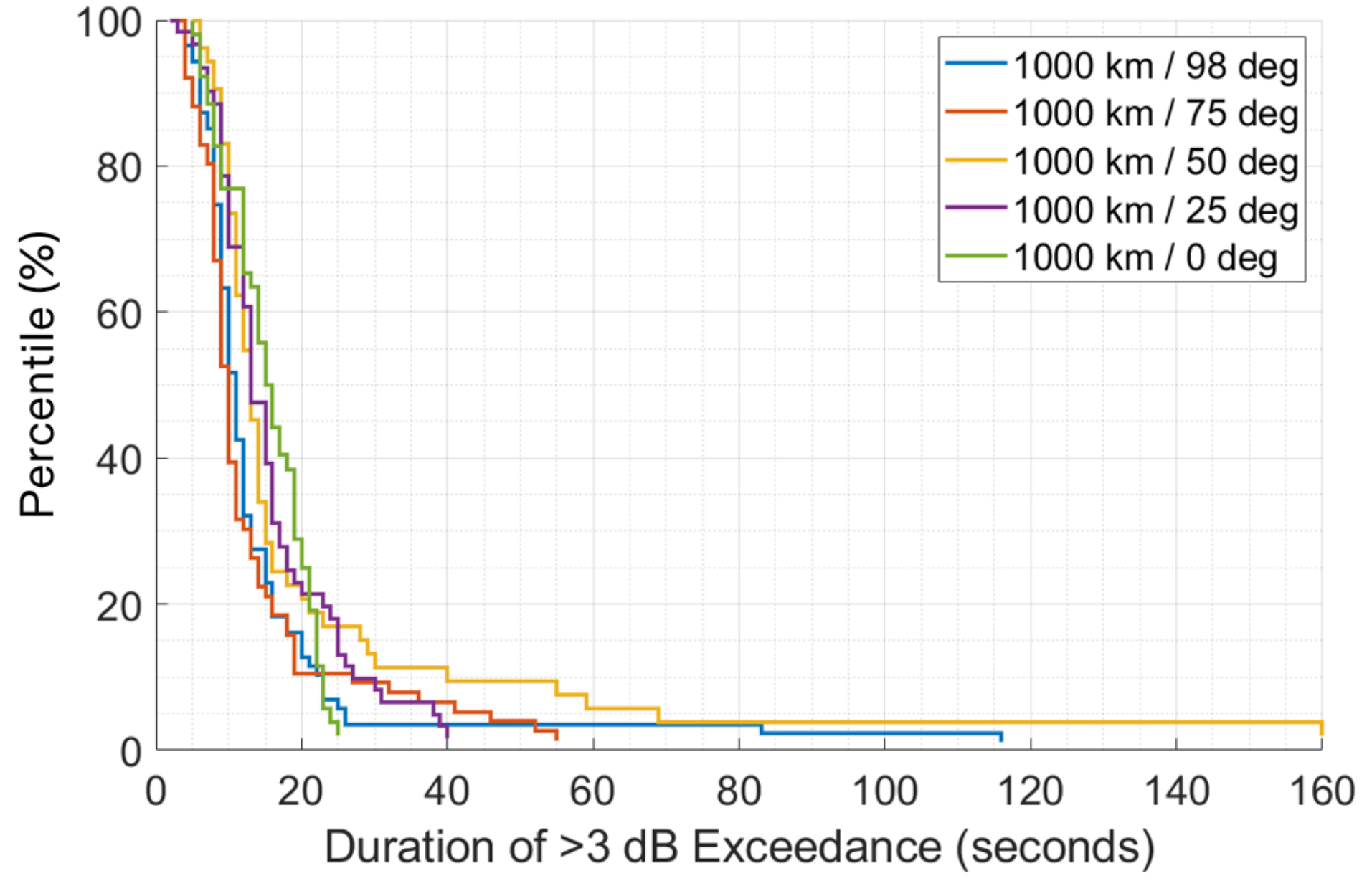

*Figure 3: Complementary cumulative distribution function (CCDF) of greater than 3 dB exceedance duration for the GPS L5 signal for satellites at 1000km orbit. Note that 99% of cases are below 3 dB, hence this shows the distribution of the remaining 1% only, highlighting the rarity of these events.*

## 5. CONCLUSION

This paper presented a compatibility assessment of the Xona Pulsar LEO PNT constellation with spaceborne GNSS receivers operating at altitudes between 400 km and 1000 km. In the absence of a dedicated methodology for evaluating RNSS compatibility in orbital environments, the established ITU-R M.1831-1 framework was adapted to account for the dynamic geometry between orbiting transmitters and receivers. Both a static worst-case analysis and a time-varying dynamic analysis were performed to evaluate the impact of Pulsar X1 and X5 signals on existing GPS and Galileo services.

The static analysis demonstrated that $C/N_0$ degradation remains limited for receiver altitudes up to 900 km and increases only for spacecraft operating in close proximity to the Pulsar orbital shell. However, the dynamic analysis showed that these worst-case conditions occur infrequently and for short durations. For worst case users operating at 1000 km altitude (80 km below the Pulsar orbital shells), GPS L1 never exceeded a $C/N_0$ degradation of 1 dB. Near GPS L5, degradation exceedances of more than 3 dB were short, generally less than 20 seconds, and occurred less than 1% of the time. These results indicate that, despite the higher received signal power levels that can occur in orbit, LEO PNT systems can coexist with existing GNSS services without causing sustained degradation of navigation performance for spaceborne users.

Beyond the specific case of the Pulsar constellation, this work establishes a practical framework for evaluating compatibility between emerging LEO PNT systems and spaceborne GNSS receivers. As the number of navigation satellites operating in LEO continues to grow, such assessments will become increasingly important for ensuring the continued interoperability and compatibility of RNSS services in both terrestrial and orbital environments.